\documentclass[twocolumn]{aastex631}

\usepackage{xcolor}
\usepackage{comment}
\usepackage{float}
\usepackage{multirow}
\usepackage{amsmath}
\usepackage{subfigure}

\begin{document}

\title{A Deep Learning Framework for TTV Characterization: A Case
Study of the Kepler-88 System}

\correspondingauthor{Gongjie Li}
\email{gongjie.li@physics.gatech.edu}

\author[0000-0002-0231-5124]{Chen Chen}
\affiliation{School of Earth and Atmospheric Sciences,
Georgia Institute of Technology,
Atlanta, GA, 30332, USA}
\email{chen_chen@gatech.edu}

\author{Jiapeng Gao}
\affiliation{Center for Relativistic Astrophysics,
School of Physics,
Georgia Institute of Technology,
Atlanta, GA 30332, USA}
\email{jgao370@gatech.edu}

\author[0000-0001-8308-0808]{Gongjie Li}
\affiliation{School of Earth and Atmospheric Sciences,
Georgia Institute of Technology,
Atlanta, GA, 30332, USA}
\affiliation{Center for Relativistic Astrophysics,
School of Physics,
Georgia Institute of Technology,
Atlanta, GA 30332, USA}
\email{gongjie.li@physics.gatech.edu}

\author{Lingkai Kong}
\affiliation{School of Mathematics,
Georgia Institute of Technology,
Atlanta, GA, 30332, USA}
\email{lkong75@gatech.edu}

\author{Molei Tao}
\affiliation{School of Mathematics,
Georgia Institute of Technology,
Atlanta, GA, 30332, USA}
\affiliation{Machine Learning Center,
Georgia Institute of Technology,
Atlanta, GA, 30332, USA}
\email{mtao@gatech.edu}

\begin{abstract}


Transit timing variation (TTV) provides rich information about the mass and orbital properties of exoplanets, which are often obtained by solving an inverse problem via Markov Chain Monte Carlo (MCMC). In this paper, we design a new data-driven approach, which potentially can be applied to problems that are hard to traditional MCMC methods, such as the case with only one planet transiting. Specifically, we use a deep learning approach to predict the parameters of non-transit companion for the single transit system with transit information (i.e., TTV, and Transit Duration Variation (TDV)) as input. This approach is constructed using the \textit{Transformer}-based architecture that can extract long-range interactions from TTV sequential data. We focus on two-planet systems based on Kepler-88 as a starting point, with the inner planet fixed to the properties of Kepler-88 and the mass and orbital elements of the outer planet varied. Applying to Kepler-88, we obtained an overall fractional error of $\lesssim$10\% on semi-major axis and eccentricity, and $\sim 10-30\%$ on planet mass (assuming that the stellar and transiting planet parameters are known), based on a training set of a million simulations. When applied to the test set, our model achieves low fractional errors of less than a few percent for systems with TTV semi-amplitudes ranging from $\sim 0.3$ s to $\sim 5$ min. Extending our approach to a broader range of planetary architectures, including systems with substantially weaker or stronger TTV signals and those containing additional planets or stellar companions, remains an important direction for future work.

\end{abstract}

\keywords{Exoplanet astronomy; Exoplanet systems; Exoplanet detection methods; Machine Learning; }


\section{Introduction} \label{sec:intro}

More than 5000 exoplanets have been discovered, and a large majority of them are detected via transit methods\footnote{\url{https://exoplanetarchive.ipac.caltech.edu/}}. Transiting exoplanets provide important information about the demographics and architectural properties of the planets \citep[e.g.,][]{Lissauer+11, Fang+13, Fabrycky14, ballard16, Weiss+18Peas}. However, observational limitations render our knowledge of many of them incomplete. For instance, even small mutual inclination between the planets may cause some of the planets to be misaligned from the line of sight and thus undetectable via transits. 

A range of dynamical mechanisms can increase planetary mutual inclination and forbid them from all transiting, such as planet-planet scattering \citep{Chatterjee+08}, secular resonances/chaos \citep{LithwickWu+11, naoz+13, Petrovich_2019, chen+22, Faridani23}. In particular, even for transiting multi-planetary systems, \citet{faridani2024more} showed that a large fraction ($\sim 20\%$) of them can undergo sweeping secular resonances that enhance their mutual inclination. These unseen planetary companions can in turn perturb orbits of the transiting planets and influence the overall dynamical evolution of the systems \citep{Becker+17, Pu+18, Denham+19, Faridani+22}.

Searching for non-detected companions of transiting planets is crucial for understanding the orbital architectural structure of the planetary systems, which will help us better understand the demographics and the formation channels of planetary systems. For transiting systems, planet-planet interactions lead to changes in their transit times, and the resulting transit timing variations serve as a powerful method to determine mass and orbital parameters of the transiting planets \citep{Dobrovolskis96, Miralda02, Holman05, Agol05}, as well as discovering non-transiting companions \citep{Ballard11, Nesvorný_2013, Saad-Olivera17}. Even for hot Jupiters which are mostly expected to be loners, \citet{Wu_2023} reported that at least 12\% $\pm$ 6\% of them, as well as the majority of warm Jupiters, host a planetary companion by analyzing TTV signatures.

\begin{figure}[ht!]
\centerline{\includegraphics[height= 2.5in]{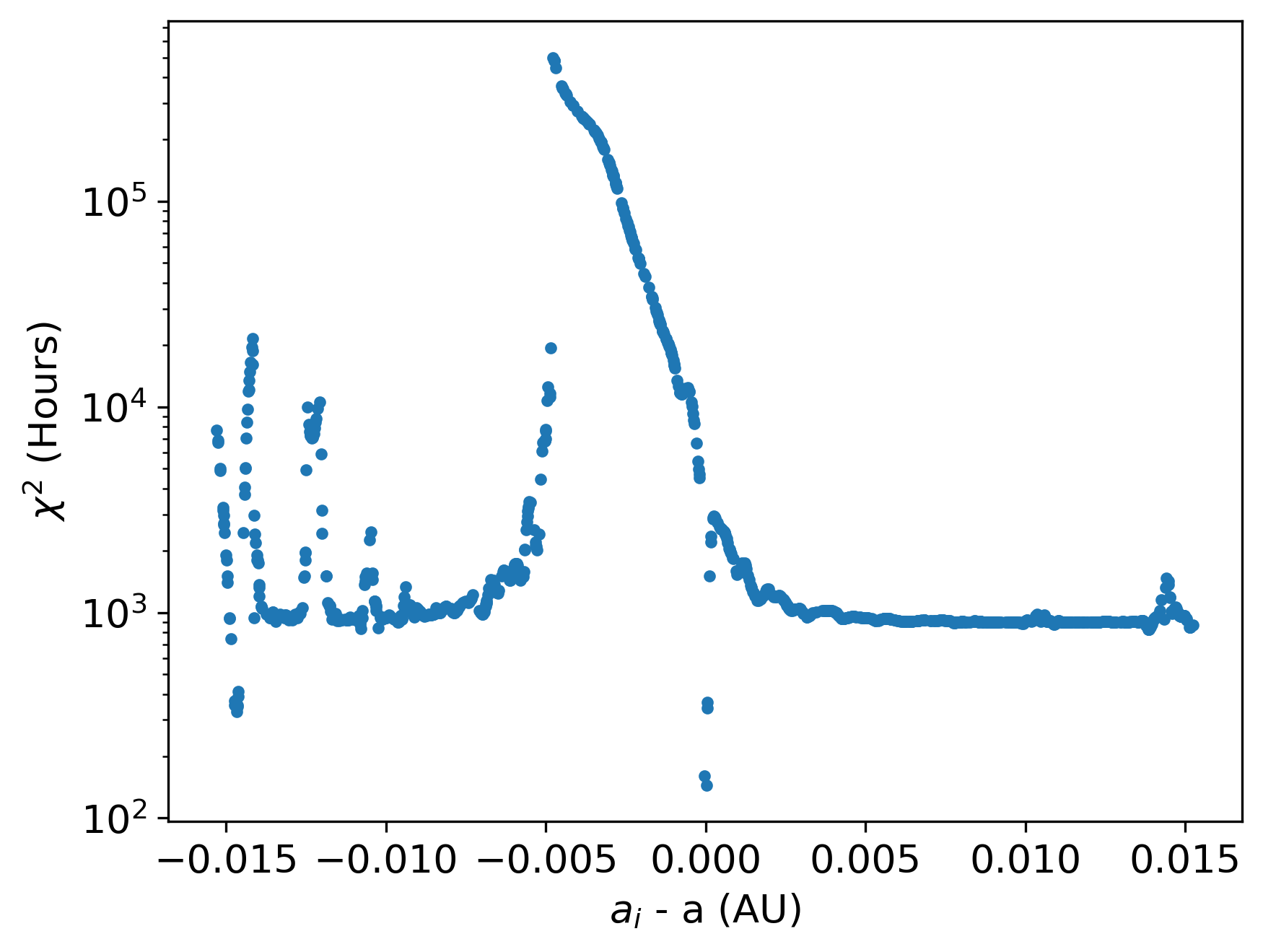}}
\caption{How TTV changes as the semi-major axis varies. Peaks demonstrate the lack of regularity (i.e. smoothness) in the response. Results obtained from high accuracy N-body simulations \citep{grit}. The system parameters are based on \cite{Nesvorný_2013}, with the semi-major axis varying within $10\%$ of the given value of 0.15292 AU (1000 different values simulated), and other parameters fixed. x-axis is the difference between the tested and given \cite{Nesvorný_2013} semi-major axis values. y-axis is $\chi^2 = \sum \frac{(V_{fixed} - V_{sim})^2}{abs(V_{sim})}$, where $V_{fixed}$ is the simulated TTV with fixed parameters based on \cite{Nesvorný_2013} while $V_{sim}$ represents the simulated TTV with varied semi-major axis.}\label{fig:resonanceCreatesJumps}
\end{figure}

  A Bayesian inverse problem approach is often used to derive the masses and orbital parameters of exoplanets from TTV \citep[e.g.,][]{Sanchis-Ojeda12, Huber13, Masuda_2017, Lam_2020}. The idea is to postulate a prior distribution on the parameters, construct a likelihood function based on the planetary dynamics that quantifies the likelihood of observing the TTV data given one set of parameter values, then correct the prior using the observations, i.e. construct a posterior distribution conditioned on the observations, and finally, sample from the posterior to statistically estimate the planetary parameters. However, this approach is computationally expensive, sensitive to the prior distribution, and not always accurate. For example, statistically speaking, the posterior distribution is often multimodal and algorithmically challenging to sample from. Moreover, the likelihood function may not be continuously differentiable, e.g., due to resonant planetary dynamics in our case (see e.g., Fig.\ref{fig:resonanceCreatesJumps}), which renders popular gradient-based samplers such as HMC/NUTS \citep{hoffman2014no} no longer applicable; affine invariant ensemble sampler \citep{goodman2010ensemble} may help, but mixing could be slow and thus the posterior sample quality may not be ideal. 

  In addition, physically speaking, when the system has only one transiting planet, the properties of non-transit planets are even harder to obtain, since the unknown planets require an even larger parameter space to fit for the TTVs. 

  On the other hand, deep neural networks demonstrated impressive performance across a wide range of application domains. In astronomy, and particularly for the detection of exoplanets, machine learning approaches have been developed to assist the search for transiting exoplanets based directly on transit light curves \citep[e.g.,][]{Shallue18, Armstrong21, Valizadegan_2022}. For TTVs, \citet{Pearson_2019} developed an ML algorithm using CNN to obtain a smart prior for the perturbing undetected exoplanet, based on the mass and orbital parameters of a known transiting planet. 
  We also note \citet{Ikhsan2024} recently extended this approach (mainly) by changing the training data set to those produced by Rebound and TTVFast. 

In this paper, we leverage the \textit{Transformer} architecture \citep{vaswani2017attention} which is more recent than CNN. Transformer can capture long-range dependence in a sequence, 
specially when the training dataset is large \citep[e.g.,][]{brown2020language}. It is therefore rather suitable to our task. More precisely, due to dynamical reasons, long-range interactions in TTV sequence provide vital information for determining planetary parameters. Moreover, our method is built such that it uses a synthetic (as opposed to observational) training set (see Sec.\ref{sec:method}), which can thus be as large as needed because it is generated by N-body simulation. 
The neural network approach is plausible, because we leverage the universal approximation capability, even for less smooth functions, of deep learning models, and indeed promises are demonstrated in finding a trained neural network model that maps TTV, as well as TTV+TDV sequence to planetary parameters.

This paper is organized as follows: In the method section (Section \ref{sec:method}), we firstly demonstrate the data preparation in Section \ref{sec:data}, and secondly introduce our model in Section \ref{sec:model} by illustrating the model architecture and giving a brief review of techniques we use, and then present the training and tuning procedures in Section \ref{sec:train}. In Section \ref{sec:res}, we show the prediction on Kepler-88 and on the testing data set we generate to illustrate our model performance. Finally, Section \ref{sec:limit} discusses the limitations and scope of applicability of our model, and Section \ref{sec:conclude} concludes the paper by summarizing the work and describing potential future directions. Kepler-88, a two-planet system with only one transiting planet, will be used as a main illustrative system throughout the paper.  

\section{Method}\label{sec:method}
In this work, we focus on two-planet systems, in which one planet transits while the other does not.
The goal is to predict the non-transit planet mass and orbital parameters (\textit{a, e, i}) with the known transit information (transit time, TTV, TDV, average duration, and average orbital period) of the transiting planet as the input for the deep learning model. 

To this end, we adopt the following work flow: we first enumerate possible values of the unknown parameters, and calculate the TTV, TDV, average duration and period of the transiting planet corresponding to each parameter set, using the high accuracy simulation package Gravitational Rigid-body Integrators (GRIT) \citep{grit} (note it works for both N-body (point masses) and N-rigid-bodies (planets with nonzero volumes)). The transit information is calculated based on the position of the planet and the size of the star. The result constitutes our training data set, which is however not large enough to contain the actual observed TTV sequence. Then we construct a specialized neural network model, by incorporating the attention mechanism from a powerful deep learning architecture known as Transformer with another popular architecture known as Gated Recurrent Unit (GRU \citep{cho2014learning}, which is a variant of Recurrent Neural Network (RNN)). 
The idea is to use GRU to encode the initial input, so that the downstream transformer component has input with better processed information; see Section \ref{sec:model} for more details as well as why this hybrid architecture.
Afterward, we train this model with transit information being the input and planetary properties being the output. 
Once the model is trained, we provide it with input being the TTV observation data (no longer synthetic), without denoising, and collect the model's output as the predicted planetary parameters. The error on the testing set (i.e., the data we prepare and not involved in the training process) is shown to demonstrate our model performance (see Section \ref{sec:res_perf}).

We discuss the details of data preparation, deep learning model, and training and tuning process in Section \ref{sec:data}-\ref{sec:train}, respectively.

\subsection{Data Preparation} \label{sec:data}

Our demonstrative example, Kepler-88, hosts 2 planets, b (transiting) and c (non-transiting). The parameters of this system have been reported in previous works(e.g., \cite{Nesvorný_2013}, \cite{Weiss_2020}). 
The parameter choices that we use in numerical simulations for preparing the data are summarized in Table \ref{tab:k88}. 

We assume that we know properties of the transiting planet (e.g., through RV observation), and we fix the parameters of planet b and the star based on \cite{Nesvorný_2013}. For non-transit outer planet c, its parameters are uniformly distributed within wide ranges as shown in Table \ref{tab:k88}. The mass is between $10 M_E$ and $10 M_J$, corresponding to $\sim 5\%$ to $\sim 16$ times that of the estimation for Kepler-88c \citep{Nesvorný_2013}. The semi-major axis ranges from 0.117 AU to 5 AU, where the lower bound of 0.117 AU is the orbital distance with a 15 days orbital period, and is close to 6 mutual Hill radius with $M_c = 10 M_E$. Closer separations may lead to orbital instability \citep[e.g.,][]{Gladman93, Volk24}, and the larger limit of the semi-major axis is more than 30 times that of Kepler-88c. We consider a low eccentricity, ranging from 0.01 to 0.1. We assume a near co-planar configuration and set the inclination up to $\sim7.7^\circ$ (around 3$\sigma$ reported in \citep{Nesvorný_2013}). The lower bound in Table \ref{tab:k88} is defined as the minimum inclination that prevents the planet c from transiting, which depends on the stellar radius and the orbital distance. Regarding the argument of periastron, longitude of ascending node, and mean anomaly, they are uniformly distributed over $0^\circ$ and $360^\circ$. The simulation runs about 4 years and stops when 135 transit epochs have been recorded. 

To calculate the transiting information in the N-body simulation, we record the times when the planet starts and ends each transit with x-axis as the reference direction. The calculation depends on the stellar radius and the projected distance on the Y-Z plane of the planet. The transit timing is in the middle of start and end times. Then, TTV, TDV, average duration and period can be derived.


The training data is then filtered to rule out sample points whose transit intervals deviate significantly from the observation. Specifically, we adopt the observed transit time from \cite{Holczer_2016} and calculate the interval between available adjacent transits. If the corresponding interval in the simulation is different from the observed one by larger than $1\%$, this simulation would be aborted and neglected. Additionally, if the transit planet does not transit or the non-transit planet transits or the current orbital distance is larger than 10 times of the initial distance (indicating instability), the simulation would be abandoned as well. As a final filtering step, we exclude the data with a TDV larger than 0.5 given that the maximum observed TDV for Kepler-88 is smaller than 0.1. 

We note that the filter is very lenient: 1\% of the transit interval corresponds to $\sim 2.6$ hr, and applying this criterion does not exclude systems with TTV semi-amplitudes smaller than a fraction of a second (see Figure \ref{fig:frac_erro_semiamp}). Overall, only 3.7\% of the simulations are filtered out. The retained sample therefore still spans a diverse range of planetary configurations in both the training and testing sets, avoiding the risks of overfitting to a single system. Indeed, as discussed in Section \ref{sec:res_perf} and \ref{sec:res_spec}, the model performs better on the test sample than on Kepler-88, indicating that it has not been tuned specifically to the Kepler-88 system.

After filtering, we obtain $\sim 1.85$ million data in total. We split $98\%$ for training purpose, $1\%$ for evaluation, and the rest $1\%$ for testing. We find the model based on the error on the evaluation set and check the performance by the error of the testing set. Additionally, we generate one more testing data and use the parameters of Kepler-88 derived from \cite{Nesvorný_2013} as initial conditions. Therefore, we can have the ground truth values to check the prediction accuracy for the system approximately. We also use the observation of Kepler-88 from \cite{Holczer_2016} as the input to make the prediction. For the missing transits in the observation, we do interpolation to make them consecutive in order to be consistent with the data we generate. We adopt Locally Weighted Scatterplot Smoothing (LOWESS) function to smooth the observed TDV with a fraction of 0.25. Each simulation is short and the simulations can run in parallel, thus, it only takes us one to two days to prepare the training data.
    
\begin{deluxetable*}{ccc}
\tablecaption{Summary of parameters of Kepler-88 System for N-body simulation \label{tab:k88}}
\tablewidth{0pt}
\tablehead{
\colhead{} & \colhead{\textbf{Planet b}} & \colhead{\textbf{Planet c}}
}
\startdata
Mass & $8.7 M_E$ & $10 M_E - 10 M_J$\\
Semi-major axis a & 0.095 AU & 0.117 - 5 \ AU\\
Eccentricity \textit{e}           & 0.05593 & 0.01 - 0.1\\
Inclination                       & $0.945^\circ$ & \text{lower bound} - $7.7^\circ$ \\
Argument of Periastron            & $-179.41^\circ$ & $0^\circ - 360^\circ$\\
Longitude of ascending node& $270^\circ$ & $0^\circ - 360^\circ$\\
Mean anomaly & $-84.185^\circ$ & $0^\circ - 360^\circ$\\
\\
\textbf{Stellar properties}\\
Mass ($M_{\rm sun}$): 0.956\\
Radius ($R_{\rm sun}$): 0.88
\enddata
\tablecomments{The parameters of transiting planet and star are based on \cite{Nesvorný_2013}. }
\end{deluxetable*}

\subsection{Deep Learning Model}\label{sec:model}

\subsubsection{Overall Model Architecture}\label{sec:model_arch}
TTVs are sequential data, and some of the most popular models designed for sequential data are Recurrent Neural Network (RNN) and Transformer. Both models have their own benefit: RNN is easier to train at least in terms of hyperparameter tuning, and less data is required. However, it may have difficulty in capturing long term dependencies in the sequence even with improvements such as Long Short-Term Memory (LSTM) and Gated Recurrent Unit (GRU). Transformer is more expressive and achieving the state of the art performance in many tasks, e.g., machine translating \citep{vaswani2017attention}. However, it is computationally expensive to train, and performs worse than less complicated models empirically when less data is provided. What's more, training Transformers can be tricky; for example, it is sensitive to warmup \citep{xiong2020layer}.
Instead of using just transformer or just RNN, we create a hybrid architecture that combines both, due to 1) relevant prior research, 2) contemporary deep learning intuition and empiricism, and 3) experimental results specific to our task. We will first describe the architecture in the following paragraph, and then elaborate these three reasons a little more.

\begin{figure*}[ht!]
\centerline{\includegraphics[height= 4 in]{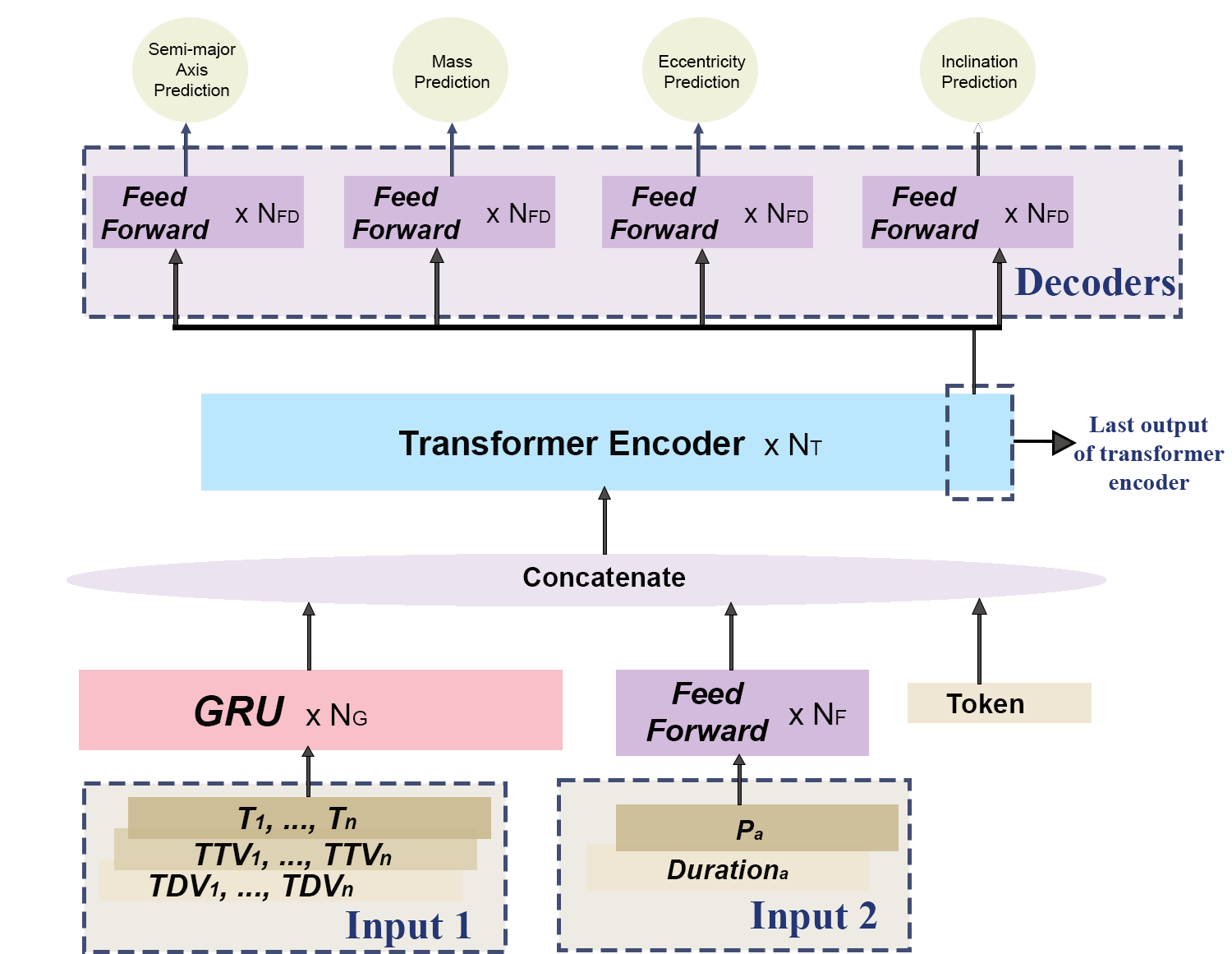}}
\caption{Deep learning model architecture. $N_G$ and $N_F$ denotes the number of layers of GRU and FF, respectively. $N_T$ is the number of layers of transformer encoder. $N_{FD}$ is the number of FF layers in decoders. Token is an extra input of the transformer encoder, which is initialized as a tensor with zeros. There are three model variants in this work: the model with TDV input and without noise, the model without TDV input and without noise, and the model without TDV input and with noise. The numbers of layers, hidden dimensions, and other hyperparameters differ among these variants, as described in Sections \ref{sec:train} and \ref{sec:res_kepler88} and implemented in the released code. As a representative example, for the model with TDV input and without noise, Input 1 has dimension $[64, 135, 3]$, corresponding to batch size, number of transit epochs, and the three input features: transit time, TTV, and TDV. Input 2 has dimension $[64, 1, 2]$. The GRU output has dimension $[64, 135, 256]$. After concatenating the GRU output, the encoded Input 2 token, and an additional zero token, the Transformer output has dimension $[64, 137, 256]$. The final output dimension is $[64, 4]$.}
\label{fig:model_arch}
\end{figure*}

Our model architecture is illustrated in Figure \ref{fig:model_arch}, which is a multi-task 
model with multiple decoders. It is mainly constructed by gated recurrent unit (GRU), feed-forward layer (FF), and transformer encoder. 
This architecture is inspired by \cite{mousavi_ellsworth_zhu_chuang_beroza_2020}, in which they constructed a deep learning model for earthquake detection and phase picking. However, we revise their architecture to better fit our problem. Specifically, CNN is not included in our model while we keep the idea of using RNN before transformer. Instead of using LSTM as in \cite{mousavi_ellsworth_zhu_chuang_beroza_2020}, we replaced it with GRU because Transformer+LSTM generally underperformed the Transformer+GRU architecture in our hyperparameter search, while also requiring slightly greater training resources and a larger model size. The way of processing output from the transformer is also different. Instead of taking the whole sequence, we only select one element from the output sequence of the transformer and then send it to the decoders, which is based on the idea from Vision Transformer \citep{vit}. The transformer encoder is then followed by feed forward layers, which serve as decoders as commonly used in scientific time-series forecasting.
The details of our model architecture are further explained in the following paragraph.

The reasons for our design of this specific hybrid architecture, as briefed above, are: 1) \cite{mousavi_ellsworth_zhu_chuang_beroza_2020} demonstrated a hybrid architecture successful for earthquake time series data, which shared similarities to the TTV data considered here.
2) Transformer is good at capturing long-range dependencies, but it has weak inductive bias for temporal continuity, is data-hungry, and can be unstable for long, noisy physical signals \citep[e.g.,][]{wen2022transformers,zeng2023transformers}. In contrast, GRU has a strong temporal inductive bias that excels in capturing short-range dependencies. Since planetary dynamics are across multiple timescales, having both local dynamics and global dependencies, a hybrid model like ours would split the job, where GRU handles local temporal evolution and transformer captures long-range/global structure. 
In fact, at the initial submission of this paper, (pure) transformer was still arguably the most popular architecture for sequential data; however, as for now (a couple of years later), the machine learning community has gradually shifted to hybrid models \citep[e.g.,][which is largely based on Mamba+transformer - note Mamba, or state space model in general, is a modern replacement of RNN (including GRU) that was not popularized yet when this paper was submitted]{blakeman2025nvidia}. 
3) We empirically tested pure transformer, transformer+LSTM (a variant of RNN), and transformer+GRU below. 

\begin{figure}[ht!]
\centerline{\includegraphics[height= 2.5in]{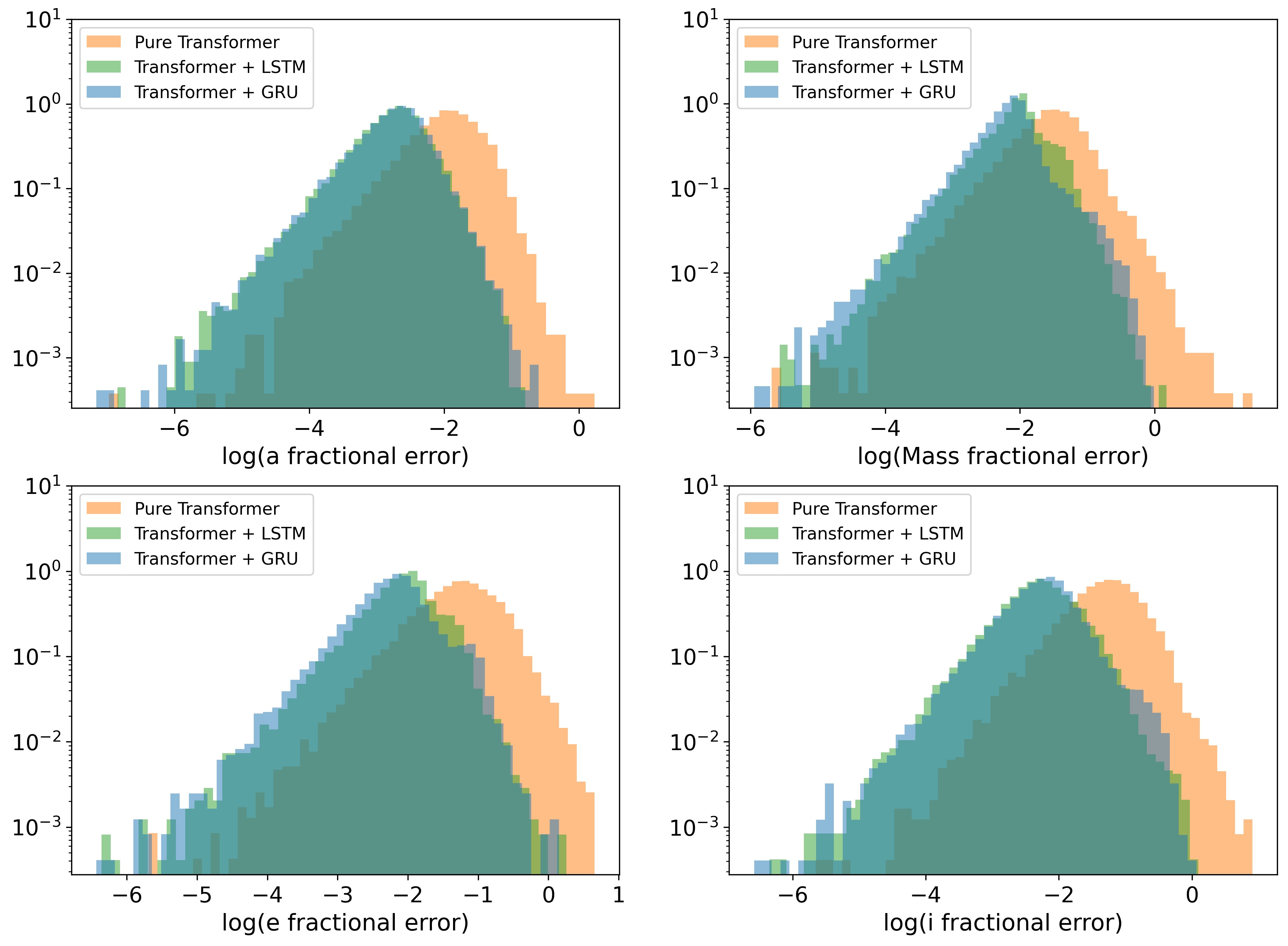}}
\caption{Fractional error distributions for predictions on the validation set using the pure Transformer, Transformer+LSTM, and Transformer+GRU architectures, shown in orange, green, and blue, respectively. The top-left panel shows the distribution for the semi-major axis ($a$), the top-right panel for the mass, the bottom-left panel for the eccentricity ($e$), and the bottom-right panel for the inclination  (i). The MSE calculated on the validation set are  $6.130\times 10^{-5}$, $6.241\times 10^{-5}$, and $2.475\times 10^{-3}$, respectively for Transformer+GRU and Transformer+LSTM and pure Transformer.}\label{fig:hist_3_architectures}
\end{figure}

The results are shown in Figure \ref{fig:hist_3_architectures}. The pure Transformer performs the worst, distinguishing it from the other two, Transformer+GRU and Transformer+LSTM. The MSE calculated on the validation set is $6.130\times 10^{-5}$ for Transformer+GRU, slightly better than LSTM which gives $6.241\times 10^{-5}$. The Transformer+LSTM result we show here is already the best-performing model after searching nearby values of hyperparameters, none of them outperforms the Transformer+GRU model. Moreover, the Transformer+LSTM architecture requires larger training resources and model size (7.32M vs 6.85M parameters) to achieve comparable performance to the Transformer+GRU architecture. Therefore, the Transformer+GRU architecture is more favorable.

As shown in the bottom of the figure \ref{fig:model_arch}, the model has two inputs. \textit{Input 1} is a two-dimensional tensor corresponding to a variable length sequence (though the length is fixed in this work) with each element being 3-dimensional, corresponding to the transit time ($T_i$), TTV, and TDV of each transit event. \textit{Input 2} has two elements, one is the average orbital period ($P_a$), and the other is the average transit duration ($Duration_a$). The GRU serves as an encoder for input 1, and FF encodes the input 2. $N_G$ and $N_F$ denote the numbers of layers for GRU and FF, respectively. The outputs from GRU and FF are concatenated sequentially with a ``token", which is an extra input initialized with zeros. This concatenated sequence is processed by the transformer encoder block, which has attention mechanism to find the relationship between each pair of elements in the sequence. $N_T$ denotes the number of layers of transformer encoder. The last element of transformer encoder output (i.e., corresponding to the ``token") is further sent to the decoders. As can be seen from Figure \ref{fig:model_arch}, each prediction has its own decoder, which is constructed by FF. $N_{FD}$ denotes the number of FF layers and it is the same in all decoders. The activation function in all FF is Leaky ReLU \citep{maas2013rectifier}. The details about our choice of the hyperparameters of the model are discussed in section \ref{sec:train}. The Pytorch framework \citep{NEURIPS2019_9015} is used to build the model.

Since the observed TDV are often noisy, we also consider the scenario that input 1 is without TDV. Under this circumstance, the model does not predict inclination, since transit timing variations alone only weakly depend on inclination, different from transit duration variations. The other parts of the model remain the same. For the model without TDV in the input, we will denote it as \textit{model without TDV} and the other as \textit{model with TDV}.

\subsubsection{Background Information: Gated Recurrent Unit (GRU) and Transformer Encoder}\label{sec:model_review}
In this section, we briefly overview the concepts of GRU and attention mechanism. More details can be found in, e.g.,  \citet{cho2014learning, vaswani2017attention}.
The GRU \citep{cho2014learning} is a type of recurrent neural network (RNN) that processes the sequential data. The basic idea of an RNN is that it reads the data sequentially and the output at $t^{th}$ step depends on the inputs before (including) $(t-1)^{th}$ step. The GRU is an improved version of the basic RNN (and a simplified version of LSTM \citep{lstm}), having gating mechanism to keep or erase previous and current information. It has reset gate ($r$), update gate ($z$), hidden state ($h$), and new hidden state ($n$), The reset gate decides how much previous information to keep, and the update gate decides if the hidden state needed to update with the new hidden state. These ``decisions" made by these gates are learned by the network. The computations of these gates and hidden states are shown by the equation \ref{eq:gru}.

\begin{equation}
\begin{split} \label{eq:gru}
    r_t &= \operatorname{sigmoid}(W_{rx}x_t + b_{rx} + W_{rh}h_{t-1} + b_{hr}) \\ 
    z_t &= \operatorname{sigmoid}(W_{zx}x_t + b_{zx} + W_{zh}h_{t-1} + b_{zh}) \\
    n_t &= \tanh(W_{in} x_t + b_{in} + r_t \odot (W_{hn}h_{t-1} + b_{hn}) )\\
    h_t &= (1-z_t) n_t + z_t h_{t-1} 
\end{split}    
\end{equation}
where $W$'s and $b$'s are weights and bias which can be learned during training, the sigmoid function is defined as $\operatorname{sigmoid}(x):=\frac{1}{1+e^{-x}}$, $x_t$ is the input at step $t$, and $\odot$ is the Hadamard product. Equation \eqref{eq:gru} gives the one-layer GRU model, with the input is a sequence $\{x_t\}_{t=1}^T$ and the output is a sequence with the same length $\{h_t\}_{t=1}^T$. This allows us to stack GRU layers to have a GRU model with $N_G$ layers.

The Transformer model is proposed by \cite{vaswani2017attention}. Here, we only adopt its encoder part. Each Transformer encoder block is a composition of attention layer, layer normalization, and feed-forward layer, with additional residual connections between adjacent layers. The attention layer characterizes the interactions between all the tokens in the sequence instead of just local interactions (such as by RNN). Each attention layer consists of multiple attention heads, which help identify multiple ways in which tokens interact. Mathematically, the output of each attention head, with $X$ being its input, is given by
\begin{equation} \label{eq:attention}
    \operatorname{Attention}(Q, K, V) = \operatorname{softmax}\left(\frac{QK^T}{\sqrt{d_k}}\right)V
\end{equation}
where $Q = X W_q$ is the query matrix, $K = X W_k $ is the key matrix, $V = X W_v$ is the value matrix, and $d_k$ is the dimension of query and key. $W_q$, $W_k$, and $W_v$ are trainable parameters of that head. Then a linear transformation of the outputs of all heads gives the output of that attention layer:
\begin{align*}
    \operatorname{MultiHeadAttention}(X)&:=\operatorname{Concat}(\operatorname{head}_1, \dots, \operatorname{head}_h) W_O\\
    \operatorname{head}_i&:=\operatorname{Attention}(X W^i_q, X W^i_k, X^i W_v)
\end{align*}
where $i$ indicates which head it is. 

\subsection{Training and Tuning}\label{sec:train}
In order to optimize the hyperparameters of our model, we conduct a grid search and choose hyperparameters that yield a relatively low validation error, defined as the model error calculated on a validation dataset that is held out from the training process. This validation dataset is independent of both the training dataset and the final evaluation dataset and is used only for hyperparameter selection, thereby preventing data leakage between model training, hyperparameter optimization, and final evaluation. We use AdamW \citep{adamW} as the optimizer for the training. The learning rate scheduler is linear schedule with warm-up, and the warm-up steps is set to 6\% of the total steps. Specifically, the learning rate first linearly increases from zero to a maximum value and then linearly decreases to nearly zero. The loss function we use is mean squared error (MSE) as it is a regression problem.
We train the models with and without TDV separately. For the model without TDV, the model does not predict inclination but still outputs semi-major axis, eccentricity, and mass. The hyperparameters we chose for both models are described in the following.

Regarding the model with TDV, the number of GRU layer ($N_G$) is 4, the number of FF for encoding \textit{Input 2} ($N_F$) is 2, the number of transformer encoder layer ($N_T$) is 3 with 4 heads, and the number of decoder layer ($N_{FD}$) is 4. The feature size is 256 for GRU and transformer encoder. For the FF encoder of \textit{Input 2}, the feature size of first layer is 128, and the second one is 256. For the decoders, the feature size of the first layer is 256, and then each layer is reduced to half of that of previous layer. 
We choose a batch size of 64 and train 200 epochs with a maximum learning rate of 0.0008. The weights of FF layers are initialized by uniform Xavier \citep{xavier}, and those of GRU layers are by orthogonal initialization. The bias of all layers is initialized as zero.

For the model without TDV, it has 3 layers of GRU ($N_G$), 2 layers of FF ($N_F$) for the encoder of \textit{Input 2}, 3 layers of transformer encoder ($N_T$) with 16 heads, and 3 layers of FF ($N_FD$) in each decoder. The feature size of GRU and transformer is 64. Similar to the model with TDV, the first layer of FF encoder for \textit{Input 2} is 32, and the second one is 64. Within the decoders, the first layer has a feature size of 64, and the subsequent layers have a feature size of half of that of their previous layer. We choose the batch size of 64 and train 80 epochs with a maximum learning rate of 0.0008. The layer initialization is the same as the model with TDV. Overall, it takes us one day to tune the hyperparameters and train our model.

\section{Results}\label{sec:res}
As described in Section \ref{sec:train}, we train two models, one includes TDV in the input and the other does not. In this section, we show the predictions and performances of both models. In Section \ref{sec:res_kepler88}, we use Kepler-88 as an illustrative example and discuss predictions in comparison to the existing understanding of this system. For investigating the robustness of our approach in face of observational noise, we consider two cases, one is trained using $\sim$ 1.8 M data without noise (i.e. clean TTV), and the other is trained using a subset (with TTV amplitude larger than the noise) of all data but perturbed by synthetic observational noise. In order to analyze the performance of our model on general planetary systems, we then show the prediction error on the testing set in Section \ref{sec:res_perf}. In Section \ref{sec:res_spec}, we select three specific systems from our testing set to further illustrate the performance.

\subsection{Prediction on Kepler-88}\label{sec:res_kepler88}
\subsubsection{Prediction Without Noise}
The results of the prediction on Kepler-88 are listed in Table \ref{tab:pred_withTDV}.
Since we do not know the actual parameters of Kepler-88, besides the observational transit information, we include a system generated by GRIT as the input where all parameters are known.
This simulated data assumes the system has a set of parameters derived from \cite{Nesvorný_2013}. By incorporating this simulated data, it helps us evaluate and compare the predictions. In the first column of table \ref{tab:pred_withTDV}, the simulated data is denoted as GRIT and the observational data is shown as Observation. In the second column, \textit{with TDV} represents the model we use has TDV in the input, while \textit{without TDV} means the model we use does not include TDV in the input. 

As can be seen from Table \ref{tab:pred_withTDV}, both models with either simulated or observational data as input all arrive similar predictions. 
For \textit{a}, both models predict $\sim$ 0.15 AU given observational data, and the predictions with simulated data also center around $\sim$ 0.15 AU. 
For \textit{mass}, the predictions of two models differ, the one with TDV gives 90.9 $M_E$ while the other without TDV gives 369.8 $M_E$. 
Regarding \textit{e}, both models predict $\sim$ 0.06 for the observational data, and the predictions on the simulated data are also close to this value.
For \textit{i}, the model predicts $\sim 3^{\circ}$ on both the observational and simulated data.

For the simulated data, since the ground truth is known, the fractional error, which is defined by equation \ref{eq:frac_err}, is calculated and shown within the parenthesis in table \ref{tab:pred_withTDV}. In general, the fractional errors are below 30\%. Specifically, for the prediction of \textit{a}, the fractional errors by both models are $\sim$ 7\%. \textit{Mass} shows a higher error, 19.6\% for the model with TDV and 29.5\% for the model without TDV. For \textit{e}, the model with TDV gives an error of 11.5\% while the model without TDV gives an error of 2.81\%. For \textit{i}, the model with TDV shows an error of 24.2\%.

\begin{equation}\label{eq:frac_err}
    \text{fractional error} = \frac{abs(Prediction - True)}{True}
\end{equation}

\begin{deluxetable*}{cccccc}[ht!]
\tablewidth{0pt}
\tablehead{
\colhead{Input} & {Model} & \colhead{\textbf{a}} & \colhead{\textbf{Mass}} & \colhead{\textbf{e}} & \colhead{\textbf{i}}
}
\startdata
GRIT & with TDV & 0.1525 AU (0.3\%) & 176.4 $M_E$ (11.2\%) & 0.0628 (11.6\%) & $4.2^{\circ}$ (9.9\%)\\
Observation & with TDV & 0.1468 AU & 68.9 $M_E$ & 0.0720 &  $2.7^{\circ}$ \\
GRIT & without TDV & 0.1419 AU (7.22\%) &  257.4 $M_E$ (29.5\%)  &  0.0547 (2.81\%)\\
Observation  &  without TDV & 0.1539 AU & 369.8 $M_E$ & 0.0625  \\
\enddata
\caption{Prediction on Kepler-88 without noise. The observation input is the transit information from \cite{Holczer_2016}. For the input from GRIT, the parameters of a, mass, e and i based on \cite{Nesvorný_2013} are 0.15292 AU, 198.8 $M_E$, 0.05628, and $3.8^{\circ}$, respectively.}\label{tab:pred_withTDV}
\end{deluxetable*}

\subsubsection{Prediction With Noise}
In this section, we consider results including noises in observational data. Assuming that the noise reflects timing measurement uncertainties induced by photometric noise and propagated through the transit fit, we draw the noise independently from a Gaussian distribution for each transit. Specifically, we select data from total dataset with a TTV amplitude larger than 1 minute, and add noise which is randomly generated from a standard normal distribution with a 1 minute standard deviation, which matches the measured per-transit timing precision of Kepler-88. The magnitude of the noise is set primarily by the host-star brightness, the transit depth, and the cadence. 

We focus on the model without TDV in the input. The training procedure and the model set-up are the same as that described in Section \ref{sec:train} but with different hyperparameters, based on the results of a grid search for the hyperparameters. We train the model for 400 epochs with a batch size of 256 and a maximum learning rate of 0.0008, and select the one with a relatively low error on the evaluation set. The selected model has 4 layers of GRU, 3 layers of transformer encoder with 8 heads, and 4 layers of decoder. The feature size of GRU and transformer encoder is 256.

Table \ref{tab:pred_withTDV_with_noise} shows predictions for Kepler-88. With input generated by GRIT, \textit{a} is 0.1454 AU, \textit{mass} is 124 $M_E$, and \textit{e} is 0.0565. The fractional errors for these parameters are 4.92\%, 37.6\%, and 0.403\%, respectively. With observational data as the input, \textit{a} is 0.1458 AU, \textit{mass} is 194.9 $M_E$, and \textit{e} is 0.041. Both inputs give similar predictions and do not deviate much from the results without considering noise (Table \ref{tab:pred_withTDV}). This is because the TTV amplitude is large $\sim 12$ hours, and this noise level only introduces a small ($\sim 0.1\%$) perturbation on the signal. 

We note that our model assumes noise sampled independently from a Gaussian distribution and has not been tested against non-Gaussian or correlated noise. In addition, the model's noise invariance is demonstrated only for the high signal to noise system (Kepler-88), where the timing uncertainties remain substantially smaller than the TTV amplitudes. Many observed systems may have considerably lower signal-to-noise ratios, for which the robustness of the method is not guaranteed. We therefore recommend the `noise-free' model for general applications, while leaving a systematic investigation of low-SNR systems and non-Gaussian, correlated noise to future work.

\begin{deluxetable*}{cccc}[ht!]
\tablewidth{0pt}
\tablehead{
\colhead{Input} & \colhead{\textbf{a}} & \colhead{\textbf{Mass}} & \colhead{\textbf{e}} 
}
\startdata
GRIT & 0.1454 AU (4.92\%) & 123.985 $M_E$ (37.6\%) & 0.0565 (0.403\%) \\
Observation  & 0.1458 AU & 194.9 $M_E$ & 0.0410 \\
\enddata
\caption{Prediction on Kepler-88 with noise. The observation input is the transit information from \cite{Holczer_2016}. For the input from GRIT, the parameters of a, mass, e and i based on \cite{Nesvorný_2013} are 0.15292 AU, 198.8 $M_E$, 0.05628, and $3.8^{\circ}$, respectively.
}
\label{tab:pred_withTDV_with_noise}
\end{deluxetable*}

\subsection{Prediction Error on Testing Set}\label{sec:res_perf}

The performance of our trained model on a single system, Kepler-88, may not represent its performance on other systems, as we train the model in a large parameter space. Therefore, in this section, we present the performance of our models by showing the fractional error on the testing set. As discussed in section \ref{sec:data}, we generate data by running N-body simulation, and the testing set is 1\% of total data generated and is not involved in the training process.

\begin{figure}[ht!]
\centerline{\includegraphics[height= 2.5in]{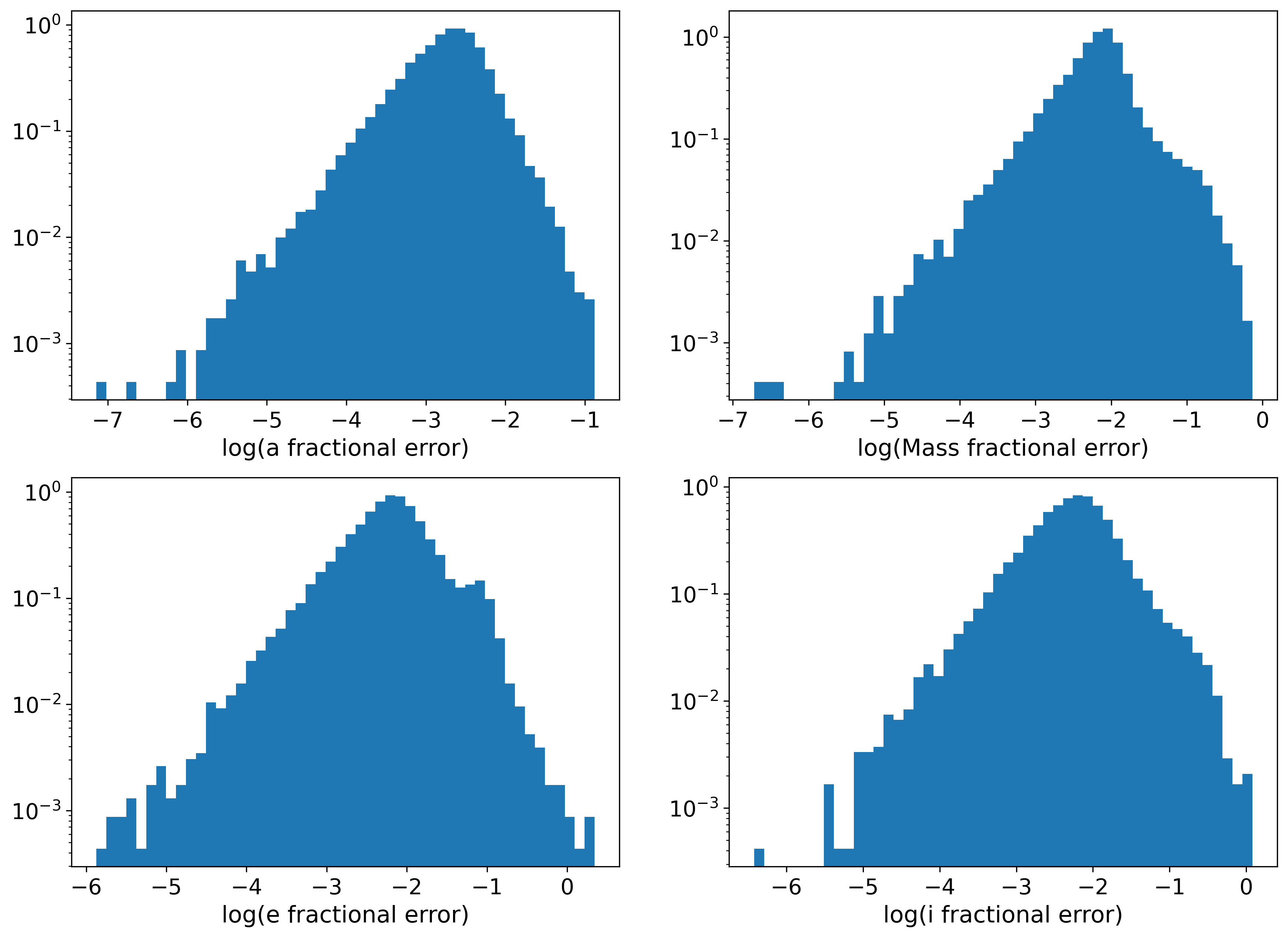}}
\caption{Fractional error distributions on the testing set prediction for the model with TDV in the input. Top left shows the distribution for semi-major axis (a). Top right shows the distribution for the mass. Bottom left shows the distribution for the eccentricity (e). Bottom right shows the the distribution for the inclination (i).\label{fig:test_err_withTDV}}
\end{figure}

\begin{figure*}[ht!]
\centerline{\includegraphics[height= 2 in]{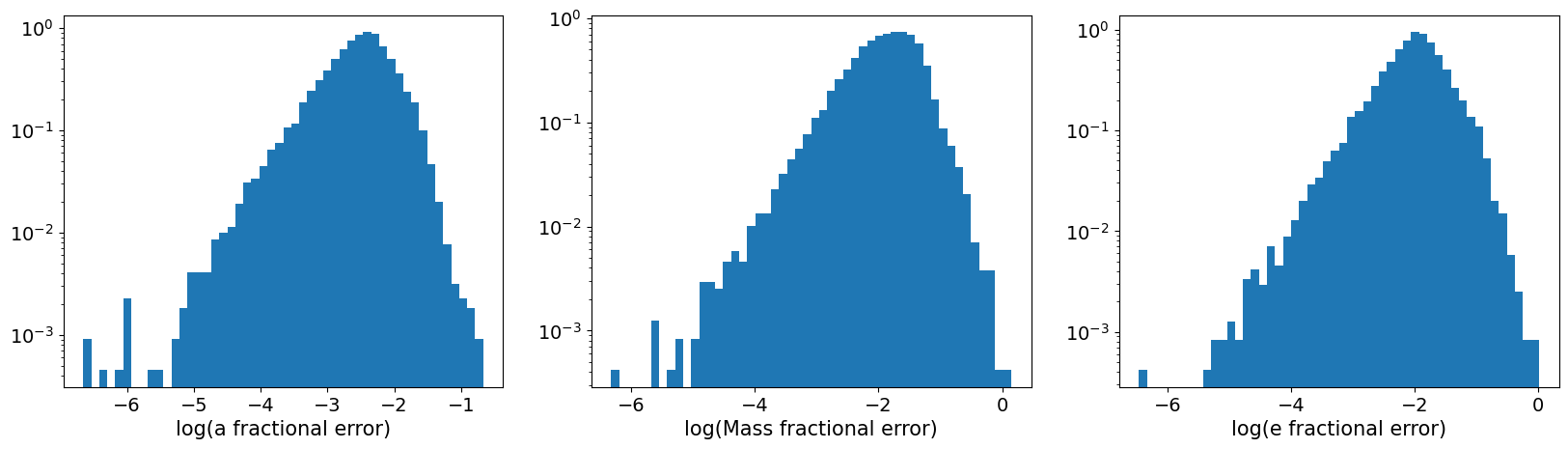}}
\caption{Fractional error distributions on the testing set for the model without TDV in the input. From left to right, it shows the distribution for semi-major axis (a), mass, and eccentricity (e), respectively.}\label{fig:test_err_withoutTDV}
\end{figure*}

Figures \ref{fig:test_err_withTDV} and \ref{fig:test_err_withoutTDV} show the fractional error distribution on the testing set. Figures \ref{fig:test_err_withTDV} refers to the model that includes TDV in the input, while figure \ref{fig:test_err_withoutTDV} is the model without TDV in the input.

Figure \ref{fig:test_err_withTDV} respectively shows the fractional error distributions for semi-major axis (a), mass, eccentricity (e), and inclination (i) of the non-transit planet. The x-axis shows the fractional error and the y-axis denotes the count, both in log scale. 
The peaks are located around $10^{-2}$ (i.e., 1\%) for all parameters, indicating a high accuracy of the predictions. More specifically, the mean (and standard deviation) of fractional errors for a, mass, e, and i are 0.31\% (0.51\%), 1.3\% (3.0\%), 1.4\% (4.1\%), and 1.4\% (4.0\%), respectively.

For figure \ref{fig:test_err_withoutTDV}, it demonstrates the fractional error distributions for semi-major axis (a), mass, and eccentricity (e). The peaks are again located near $10^{-2}$. The mean (and standard deviation) of fractional errors for a, mass, and e are 0.51\% (0.71\%), 2.4\% (3.8\%), and 1.9\% (3.2\%), respectively.

\begin{figure*}[ht!]
    \centering
    \includegraphics[height= 3.5 in]{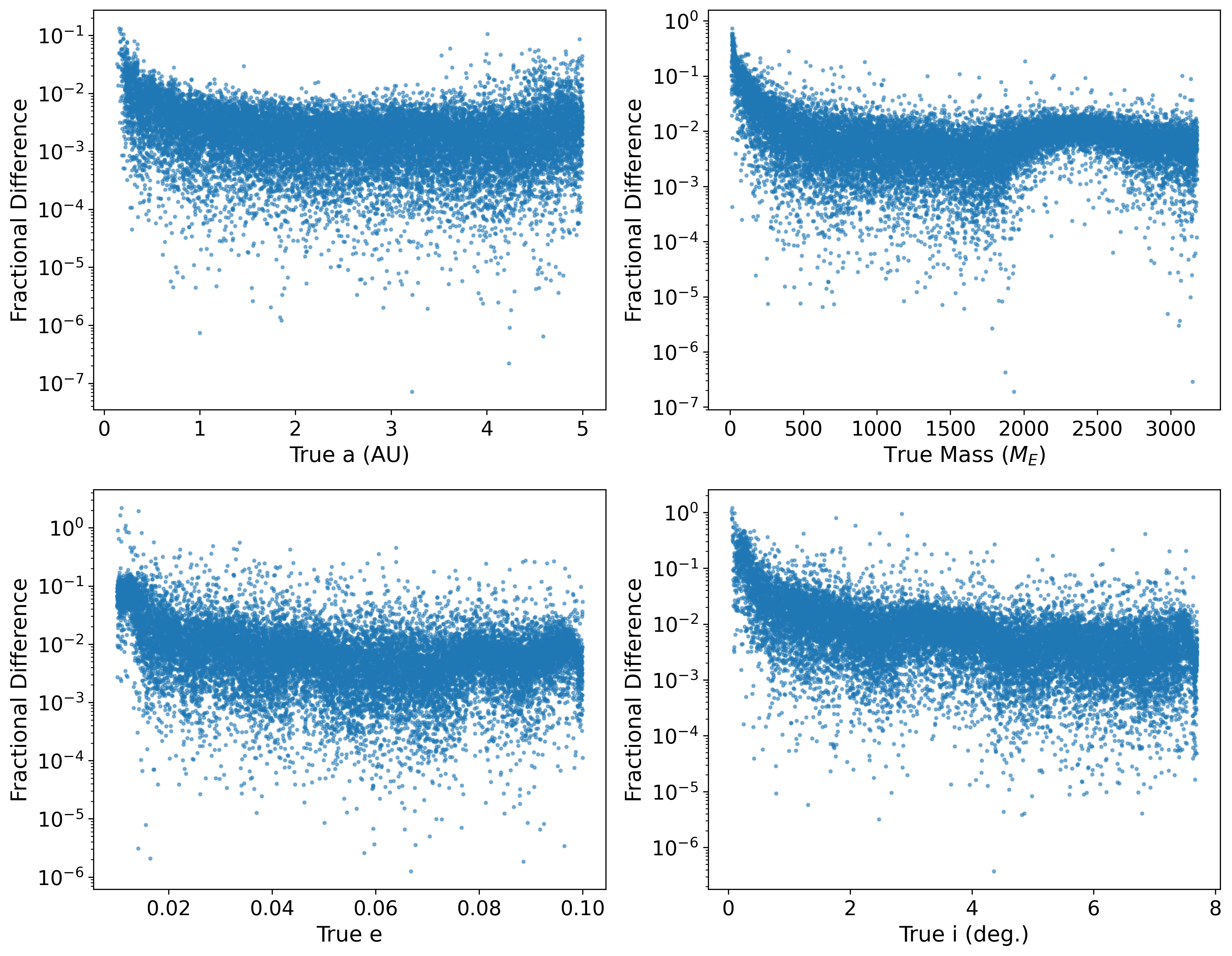}
    \caption{The relationship between the true value and the fractional error by the model with TDV. Each dot shows the fractional error for a specific system.}
    \label{fig:mad_withTDV}
\end{figure*}

\begin{figure*}[ht!]
    \centering
    \includegraphics[height= 1.9 in]{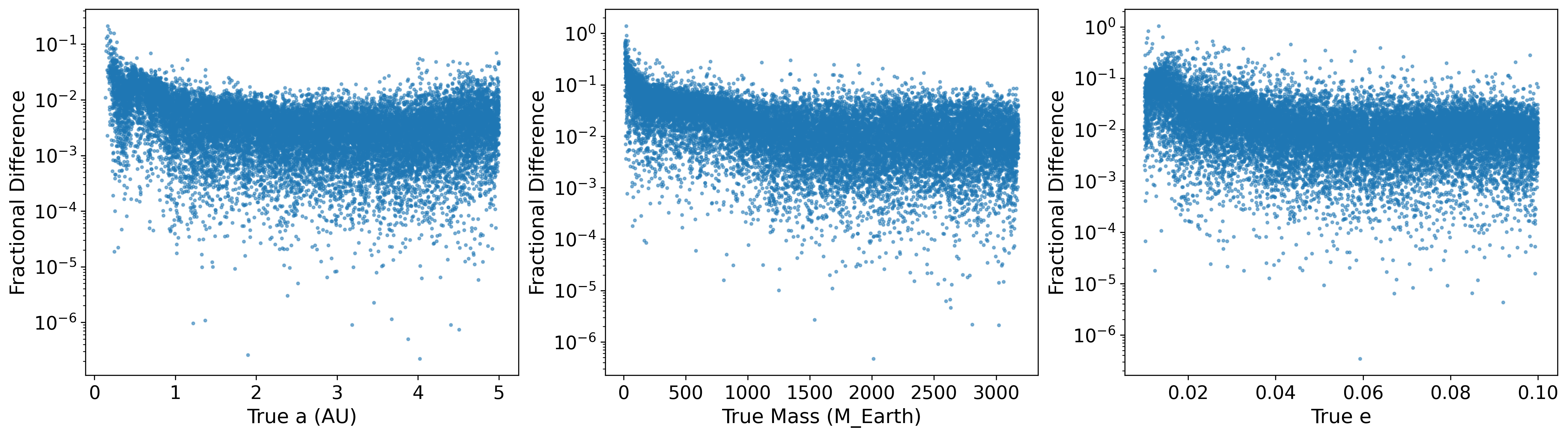}
    \caption{The relationship between the true value and the fractional error by the model without TDV. Each dot shows the fractional error for a specific system.}
    \label{fig:mad_withoutTDV}
\end{figure*}

Both models show low fractional errors in general and the model with TDV shows a slightly better performance. 
Figures \ref{fig:mad_withTDV} and \ref{fig:mad_withoutTDV} show the relationship between true/initial parameters and the fractional error for both models. The solid line represents the median fractional error and the shadow denotes the median absolute deviation. Most predictions have a fractional error of $\lesssim$1\% while a smaller value has a higher fractional error. 
This is because we use mean square error loss function during the training. The absolute error, i.e., the difference between the predicted and the true value, is similar between different systems, and thus it is natural to observe that the fractional error is larger for systems with low semi-major axis, mass, eccentricity, and inclination.
However, our model generally demonstrates the robustness and effectiveness in predicting the parameters of non-transit planet in the testing set. 


Compared to the fractional error for Kepler-88 listed in Tables \ref{tab:pred_withTDV} and \ref{tab:pred_withTDV_with_noise}, the model performs better for systems in the testing set. Changing the splitting to 80/10/10 and replacing MSE with Mean Absolute Percentage Error (MAPE) or the symmetric Mean Absolute Percentage Error do not alleviate the performance on Kepler-88. This is because our model performs worse when the true parameters lie near the lower boundary of the training range, such as the case for Kepler-88 (see more discussions in the next section \ref{sec:res_spec}). In addition, the training set consists primarily of systems with randomly generated orbital parameters for the outer perturber, which are mostly not in mean-motion resonances, unlike Kepler-88. Consequently, the model is not specifically optimized for systems resembling Kepler-88. Despite the less accurate prediction of the outer planet's mass in Kepler-88, the model successfully recovers the Kepler 88-c's orbital parameters with $\lesssim 10\%$ fractional error on semi-major axis, eccentricity, and inclination. This suggests that our method is actually applicable and performs well across a range of planetary systems. 

\begin{figure*}[ht!]
    \centering
    \includegraphics[height= 3.5 in]{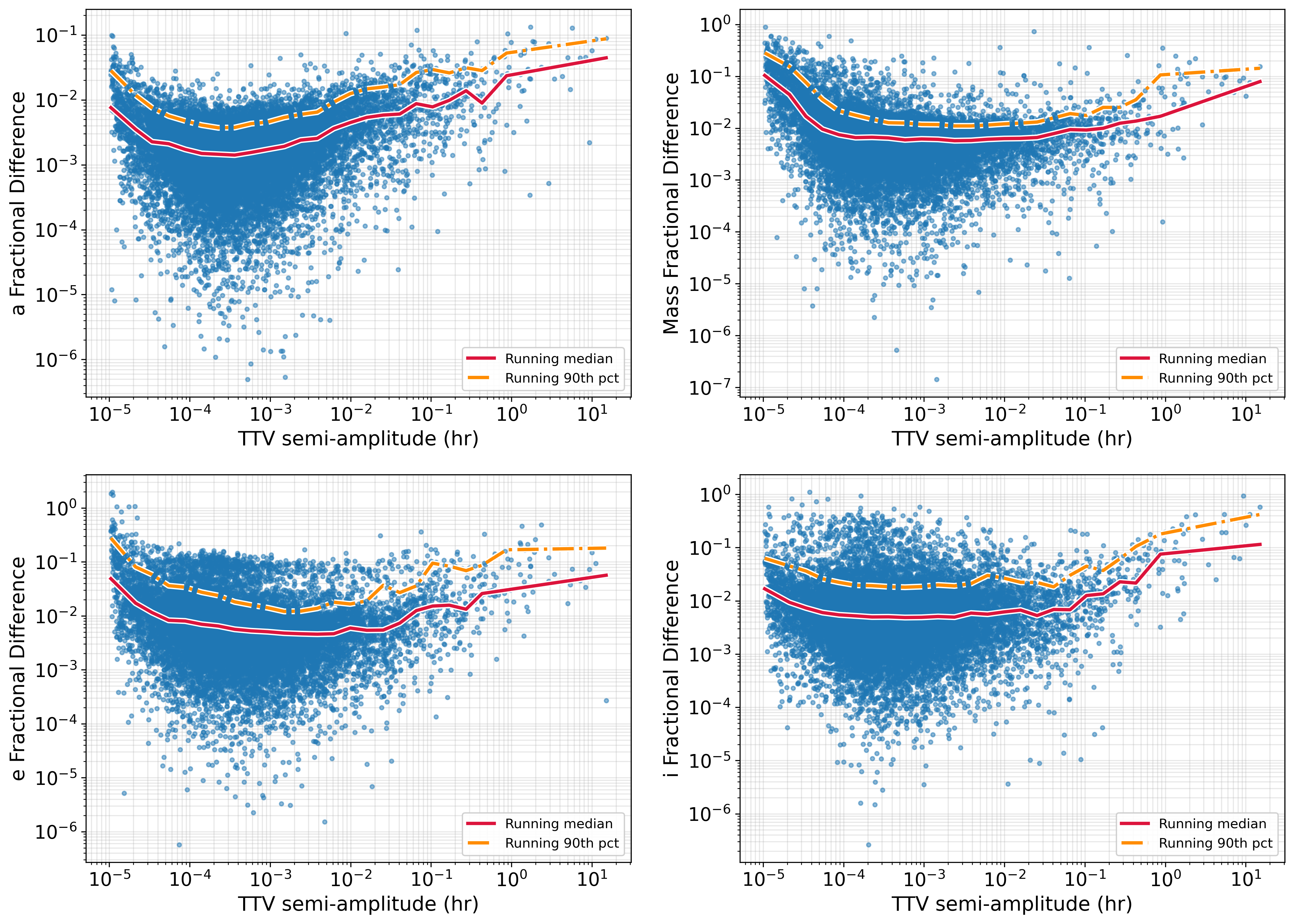}
    \caption{The fractional error versus the TTV semi-amplitude. Each dot shows the fractional error for a specific system. The orange dash-dotted line shows the running $90^{th}$ percentile, and the red line shows the running median. The fractional error are typically low $\lesssim$ a few percent for a wide range of TTV semi-amplitude of $\sim 10^{-4}-0.1$hr.}
    \label{fig:frac_erro_semiamp}
\end{figure*}

To analyze the applicability of our model more quantitatively, we plot the fractional error as a function of the TTV semi-amplitude for the model trained on TTVs without noise in Figure \ref{fig:frac_erro_semiamp}. The fractional errors generally decrease for all parameters as the TTV semi-amplitude increases from $10^{-5}$ hr to $\sim 2\times10^{-4}$ hr. This trend is expected, as larger TTV amplitudes correspond to stronger signals, allowing the model to achieve better performance.

At larger TTV semi-amplitudes ($\gtrsim 2\times10^{-4}$hr), however, the fractional error in the semi-major axis begins to increase. This behavior primarily arises because $a$ is strongly correlated with the TTV semi-amplitude: closer outer companions (smaller $a$) produce larger TTV amplitudes but also exhibit larger fractional errors, because the fractional error increases toward smaller parameter values. For the other parameters, the fractional errors remain relatively low until the TTV semi-amplitude reaches approximately $0.1$ hr, beyond which they increase significantly. This is primarily due to the limited number of training samples with TTV semi-amplitudes larger than $0.1$ hr, which constitute only approximately $1.17\%$ of the total training dataset. 
Overall, the fractional error remains below a few percent for a large range of TTV semi-amplitude between $1\times10^{-4}$ hr and $1\times10^{-1}$ hr, where the model's inference is reliable.

\subsection{Prediction on Other Example Systems}\label{sec:res_spec}

To illustrate the model performance further, we select three representative systems from the testing set, which are summarized in tables \ref{tab:example_withTDV_epoch200} (with TDV) and \ref{tab:example_withoutTDV} (without TDV). 
Two of these systems hold the non-transit planet at a further orbital distance of $\sim 3$ AU, and the masses are respectively $\sim 1 M_J$ and $\sim 5 M_J$. The third system has a much closer-in non-transit planet with an orbital distance of $\sim 0.2$ AU, and it is less massive with only $\sim 20 M_E$. 

As shown in table \ref{tab:example_withTDV_epoch200}, the model with TDV makes predictions with a fractional error (FE) of $\sim 1\%$ for systems when the non-transit planet is further and massive  (except the mass prediction for 1 $M_J$ with an FE of $4.5\%$). For the system with a close-in and lower mass non-transit planet, the FE goes up (e.g., $\sim 13\%$ for a prediction) as the values of a and mass of this system are small with respect to the parameter ranges, i.e., $\sim 0.1$ AU - 5 AU for a and 10 $M_E$ - 10 $M_J$. 
However, these FEs are still low and not significant enough to undermine our model's overall performance.

The model without TDV performs similar on these three systems as shown in table \ref{tab:example_withoutTDV}. For the systems with a $\sim 3$ AU non-transit planet, the model predicts the parameters with a low FE $\sim 1\%$. While for the system with a $\sim$ 0.2 AU and $\sim$ 20 $M_E$ non-transit planet, the FEs go up significantly for a and mass predictions. The FE for a is $\sim$ 18 \%, which is still acceptable and considered low. While for the mass prediction, the FE goes up to $\sim 71 \%$ for the system with lower planet mass $20M_\oplus$ and semimajor axis close to that of the transiting planet. However, the prediction is still within the same order and such large FE constitute only a small proportion of the testing set as shown in the Figure \ref{fig:test_err_withoutTDV}.

\begin{table*}[ht!]
    \centering
    \begin{tabular}{|c|c|c|c|c|c|c|}
        \hline
        \multicolumn{2}{|c|}{} & \textbf{$\sim$ 3AU, $\sim$ 1 $M_J$} & \textbf{$\sim$ 3AU, $\sim$ 5 $M_J$} & \textbf{$\sim$ 0.2 AU, $\sim$ 20 $M_E$}\\
        \hline
        \multirow{3}{*}{\textbf{a}} & T & 3.0255 AU & 2.9932 AU & 0.1782 AU\\
        \cline{2-5}
        & P & 3.0155 AU & 2.9770 AU & 0.1971 AU\\
        \cline{2-5}
        & FE& 0.3305\% & 0.5441\% & 10.5684\%\\
        \hline
        \hline
        \multirow{3}{*}{\textbf{Mass}} & T &  0.9924 $M_J$ & 4.9825 $M_J$ & 17.7183 $M_E$\\
        \cline{2-5}
        & P & 0.9901 $M_J$ & 4.9770 $M_J$ & 24.0963 $M_E$\\
        \cline{2-5}
        & FE & 0.2348\% & 0.1111\% & 35.9967\% \\
        \hline
        \hline
        \multirow{3}{*}{\textbf{e}} & T &  0.0746 & 0.0462 & 0.0381\\
        \cline{2-5}
        & P & 0.0752 & 0.0456 & 0.0436\\
        \cline{2-5}
        & FE & 0.7753\% & 1.2204\% & 14.5478\% \\
        \hline
        \hline
        \multirow{3}{*}{\textbf{i}} & T & 6.5460$^{\circ}$ & 6.0409$^{\circ}$ & 7.3501$^{\circ}$ \\
        \cline{2-5}
        & P &  6.4828$^{\circ}$ & 6.0620$^{\circ}$ & 7.2333$^{\circ}$\\
        \cline{2-5}
        & FE &  0.9654\% & 0.3495\% & 1.5892\%\\
        \hline
    \end{tabular}
    \caption{Prediction on selected example systems from testing set. T denotes true value, P represents prediction, and FE is the abbreviation of fractional error.}
    \label{tab:example_withTDV_epoch200}
\end{table*}

\begin{table*}[ht!]
    \centering
    \begin{tabular}{|c|c|c|c|c|c|c|}
        \hline
        \multicolumn{2}{|c|}{} & \textbf{$\sim$ 3AU, $\sim$ 1 $M_J$} & \textbf{$\sim$ 3AU, $\sim$ 5 $M_J$} & \textbf{$\sim$ 0.2 AU, $\sim$ 20 $M_E$}\\
        \hline
        \multirow{3}{*}{\textbf{a}} & T & 3.0255 AU & 2.9932 AU & 0.1782 AU\\
        \cline{2-5}
        & P & 3.0263 AU & 3.0044 AU & 0.2111 AU\\
        \cline{2-5}
        & FE& 0.0277\% & 0.3742\% & 18.4295\%\\
        \hline
        \hline
        \multirow{3}{*}{\textbf{Mass}} & T & 0.9924 $M_J$ & 4.9825 $M_J$ & 17.7183 $M_E$\\
        \cline{2-5}
        & P & 1.0081 $M_J$ & 4.9337 $M_J$ &  30.3714 
 $M_E$\\
        \cline{2-5}
        & FE & 1.5827\% & 0.9792\% & 71.4125\% \\
        \hline
        \hline
        \multirow{3}{*}{\textbf{e}} & T & 0.0746 & 0.0462 & 0.0381\\
        \cline{2-5}
        & P & 0.0749 & 0.0468 & 0.0370\\
        \cline{2-5}
        & FE & 0.3172\% & 1.2939\% & 2.7071\% \\
        \hline
    \end{tabular}
    \caption{Prediction on selected example systems from testing set. T denotes true value, P represents prediction, and FE is the abbreviation of fractional error.}
    \label{tab:example_withoutTDV}
\end{table*}

\section{Limitations and Scope of Applicability}
\label{sec:limit}
In this work, we use Kepler-88 as a representative example. Specifically, we fix the inner planet to have the same properties as Kepler-88b, a Neptune-sized planet at $\sim 0.1$ AU from a $\sim 1M_\odot$ star, while varying the properties of the outer planet, with its mass ranging from approximately a Neptune mass to $10,M_{\rm J}$, and its semi-major axis spanning $\sim 0.1$–$5$ AU (see Table \ref{tab:k88}). The overall dynamics of the planetary system can be scaled with the planetary masses and semi-major-axis separations, and this range of outer planet properties is broad. Thus, the method can be well applied to a wide range of systems. Applying to our testing sample, we find that the fractional error of our prediction generally lies below a few percent for systems with TTV semi-amplitude between $10^{-4}-0.1$ hr ($\sim 0.3$sec to $\sim 5$min).

Nevertheless, there are limitations on the applicability of our method. Our analysis assumes a two-planet system orbiting a single star, whereas additional planets or stellar companions can produce more complex TTV signatures. In addition, because we draw the outer planet's parameters from uniform distributions, our training set does not specifically target systems in or near mean-motion resonances. Consequently, the applicability of our method to planetary systems with architectures substantially different from those considered here is not guaranteed.

Accounting for the full range of possible planetary architectures, including systems with additional planets, planets in or near mean-motion resonances, or stellar companions, would substantially expand the parameter space and make both the system modeling and the generation of training data sets computationally challenging. We therefore restrict our analysis to two-planet systems in this paper with TTV semi-amplitude of $\sim 0.3$sec to $\sim 5$min. Future studies could build on our framework, potentially using transformer-based models, by extending the training set to include a broader range of planetary architectures or focusing on specific targeted systems across different dynamical regimes.

\section{Conclusion and Discussion}\label{sec:conclude}
In this work, we develop a deep learning model that predicts the parameters (a, mass, e and i) of non-transit planet directly from transit information
. We focus on two-planet systems with one planet transiting, although there is no fundamental difficulty in extending the method to cases with more transiting and/or non-transiting (i.e. hidden) planets. A significant part of our demonstration is based on Kepler-88. The predictions on Kepler-88 give a fractional error of 6.5\% for \textit{a}, 19.6\% for \textit{mass}, 11.5\% for \textit{e} and 24.2\% for \textit{i}, when the model has TDV in the input. The model without TDV has a slightly larger fractional errors on the predictions of \textit{a} (7.22\%) and \textit{mass} (29.5\%) while a lower fractional error on \textit{e} (2.81\%). On the testing set, our model shows an accuracy with fractional errors less than a few percent for most ($\sim 90\%$) samples with TTV semi-amplitude between $\sim 0.3$sec and $\sim 5$min.
To further illustrate the performance, three specific examples are selected and shown an overall small fractional error of $\sim$1\% when the outer companion is massive (1 and 5 $M_J$) at a further orbital distance of 3 AU. When the outer companion is less massive (20 $M_E$) and much closer to the host star (0.2 AU), the fractional errors are below $\sim$ 10\% but still considered low.


When the non-transit planet has a further orbital distance and higher mass, the fractional error almost always stays low. However, when the orbital distance is much closer, and mass is lower, the fractional error goes up. It is natural to have this trend using MSE loss function in our training, which leads to similar absolute error across different systems and causes the fractional error to be larger for systems with lower true values. In addition, the dynamics are more complex when the outer planet is closer, making the training harder. Moreover, we note that there are more systems that allow the non-transiting planet to transit when the non-transiting planet is located closer to the host star. Thus, more systems are filtered out when we prepare for the the training sample, and the corresponding sample size is in turn smaller for closer-in non-transiting planets. This also makes it harder to train the model and leads to larger error for non-transiting planets that are located closer to the host star.

Due to these facts, if the system has small values of parameters, it could lead to large error with respect to its true value. To mitigate this situation, several ways could address it.
Firstly, we can generate more data to train our model and (or) desample the data to make it more balanced on the distribution over \textit{a}.
Alternatively, we can narrow down the parameter space given previous results. For instance, if initially we use a large parameter space but the models consistently give a small value, then we could prepare a new training set with a smaller range to fine-tune the model.
Additionally, we could also log-scale the parameters which have a large range, and thus the model would pay more attention to the small value.

One note of the current work is that we fix the parameters of the transit planet for simplicity. In future work, we could try to vary the parameters of transit planet in a small range when preparing the training data set to potentially boost adaptability and enhance accuracy.

We note that different from the traditional MCMC approach that works on individual systems separately, the machine learning approach can be applied to a large number of systems altogether. Moreover, the performance of our method is evaluated via a statistical estimate of the overall accuracy of our predictions for an ensemble of systems. It is a limitation that it is unclear yet, as for most deep learning based approaches, how to obtain a guarantee on prediction accuracy for a specific system. Thus, different from the MCMC approach, our model does not provide a posterior distribution of the parameters. Instead, our model can obtain an initial guess on the hidden planet parameters, which can be useful for following up MCMC analysis.

The code for the results in this paper is available at \url{https://github.com/tyy-cc/DeepTTV}, and a frozen version is place at \url{https://doi.org/10.5281/zenodo.14227316}.

\section*{Acknowledgment}
The authors are deeply grateful to Judith Korth,  Hannu Parviainen, Lauren Weiss, Songhu Wang and Chelsea Huang for inspiring discussions. CC and GL are partially supported by NASA 80NSSC20K0641 and 80NSSC20K0522. LK and MT are grateful for partial supports by NSF Grant DMS-2513699 (LK \& MT), DOE Grant SC0026274 (MT), Richard Duke Fellowship (LK \& MT), and Simons Institute for the Theory of Computing at UC Berkeley (MT).
This work used the Hive cluster, which is supported by the National Science Foundation under grant number 1828187.  This research was supported in part through research cyberinfrastrucutre resources and services provided by the Partnership for an Advanced Computing Environment (PACE) at the Georgia Institute of Technology, Atlanta, Georgia, USA.
This work also utilized the Delta system at the National Center for Supercomputing Applications through allocation PHY220148 from the Advanced Cyberinfrastructure Coordination Ecosystem: Services \& Support (ACCESS) program, which is supported by National Science Foundation grants \#2138259, \#2138286, \#2138307, \#2137603, and \#2138296.


\bibliography{sample631}{}
\bibliographystyle{aasjournal}



\end{document}